# An End to End Edge to Cloud Data and Analytics Strategy

Vijay Kumar Butte[1], Sujata Butte[*2]

[*]*University of Idaho, Idaho Falls, ID, USA*
[2]suja2591@vandals.uidaho.edu
[1]vkbutte@gmail.com

*Abstract*— There is an exponential growth of connected Internet of Things (IoT) devices. These have given rise to applications that rely on real time data to make critical decisions quickly. Enterprises today are adopting cloud at a rapid pace. There is a critical need to develop secure and efficient strategy and architectures to best leverage capabilities of cloud and edge assets. This paper provides an end to end secure edge to cloud data and analytics strategy. To enable real life implementation, the paper provides reference architectures for device layer, edge layer and cloud layer.

## I. Introduction

The industries across verticals are making a tectonic shift towards cloud migration. Enterprises are moving their digital assets, services, databases and Information Technology (IT) assets to the cloud at a rapid pace. Cloud adoption enables businesses to reduce total cost of ownership, time to market and increase pace of innovation. Cloud provides improved scalability, performance and flexibility. This enables businesses to quickly support large workloads and improve their customers' experience without requiring purchase of physical servers, network equipment or software licenses. The area of cloud computing research is well established both from a technological point of view and a business point of view. The notable work in the area of compute capability, scalability, principles and paradigms can be obtained from [1-5].

Simultaneously, the world today is also witnessing enormous growth in the number of connected Internet of Things (IoT) devices. IoT systems enable connection of devices through usage of the internet and other communication technologies without needing human intervention [6]. This has given rise to many valuable applications that use the data generated by these connected devices and provide real time inferences [7]. These inferences enable businesses and individuals to identify opportunities, issues and take actions to reduce costs and increase profits. They have found valuable applications in various industries such as autonomous driving, remote health monitoring, equipment monitoring and control in industry 4.0.

The real time applications play a critical role in decision making. Hence, it becomes critical to minimize latency and enable faster processing. This gives rise to the need to bring the computation and data storage closer to the devices to overcome latency issues and improve application performance. Another aspect of bringing computation and data close to the device is to reduce overall application risk by minimizing the dependency on far away servers. This also reduces the costs associated with frequent data transfer to far away servers. This is the paradigm of edge computing where storage and computations are brought closer to devices with the aim of improving response times and reducing bandwidth costs [8-10].

The number of connected devices and the volume of data being generated by these sensors is large and increasing daily. This demands high computational power and storage capacity to manage this data. An effective way to store, analyze and consume is by leveraging cloud services. Cloud services provide the ability to store, model and process this data in scalable and cost effective manner.

For modern enterprises both edge assets and cloud assets are critical. The edge assets use data and local compute capabilities to provide quick and valuable inferences. The cloud assets enable storage, computation and advanced analytics on large amounts of data sent by these devices and other data sources. So, enterprises need a secure end to end data and advanced analytics strategy to leverage both edge and cloud assets effectively.

Latency is a key aspect for critical applications that depend on real time access to data and computational power to derive inferences. The edge service is leveraged to address this component. While cloud service is leveraged to store, process and analyze large amounts of data. This hybrid approach will enable businesses to deliver faster service efficiently. This paper develops such an end to end hybrid edge to cloud data and analytics strategy.

The core components of a hybrid cloud edge system can be classified into three layers. They are device layer, edge layer and cloud layer as shown Fig. 1.

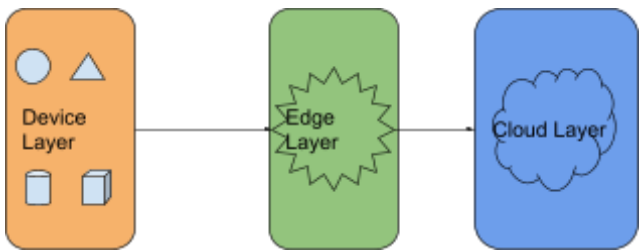

Fig.1 Core components of edge to cloud systems

In the subsequent sections, the paper dives deep into each of these layers and provides secure and efficient reference architectures. For the benefit of industry practitioners and real life applications, the Amazon Web Services (AWS) cloud versions of each of these reference architectures are provided.

## II. THE DEVICE LAYER

The device layer of the edge-cloud system includes IoT sensors, actuators, cameras, microphone etc,. They interact with the environment and translate physical reality into data that can be used to make inferences. The data generated may include information about temperature, pressure, humidity, flow, location etc. The actuators have the ability to interact with the real system based on instructions, commands or manual interventions. For example, an actuator may be used to shut off valves if the pressure exceeds a certain threshold. The IoT devices typically consist of integrated CPU, network adapter and firmware.

## III. THE EDGE LAYER

The aim of the edge layer is to minimize the latency in decision making and reduce cost. The edge component is placed close to the data source to minimize the time delay.

The key components of edge layer include edge gateway, edge event processor, edge application and edge machine learning (ML) inference. Fig. 2 shows the architectural components of the edge layer. The edge gateway connects to devices and provides a network entry point to the cloud. It facilitates bidirectional communication between devices and the cloud. The edge gateway may also provide the edge cluster. The edge cluster provides the short term storage and computation closer to the applications with the objective of reducing latency. It reduces the cost by reducing the cloud network load and reducing data transmission costs. It also facilitates preliminary operations on data. These may include data aggregations, quality check and filtering as needed by business.

The edge ML inference component performs advanced analytics, pre-processing, inference calculations close to the data source. This enables quicker decisions and needed interventions.

The edge applications consume the data from the device layer, process it and provide insights, dashboards and facilitate business actions. The applications may leverage ML inference results or act based on defined rules. The edge event processor acts based on a defined event and analyzes and takes needed actions.

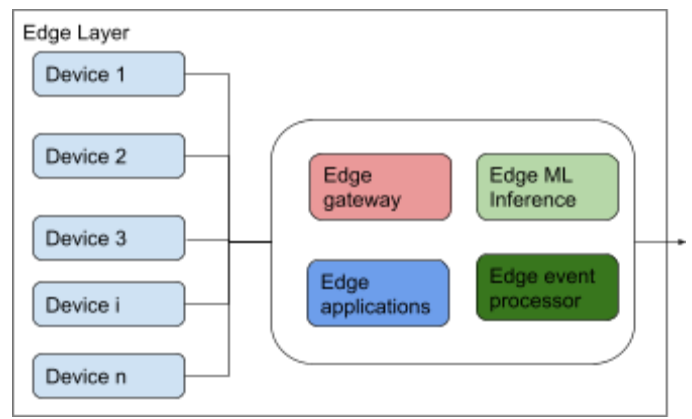

Fig. 2 Architectural components of edge layer

Amazon IoT Greengrass enables data collection and analysis close to the data source. It establishes secure connection between devices and acts based on the edge events. The IoT Greengrass service also helps to manage edge device softwares. The AWS Lambda facilitates execution of edge applications based on defined triggers. The machine learning model development takes place in the cloud where the needed data, compute, networking and storage infrastructure is available. The machine learning model is deployed at the edge through Amazon SageMaker for real time inferences.

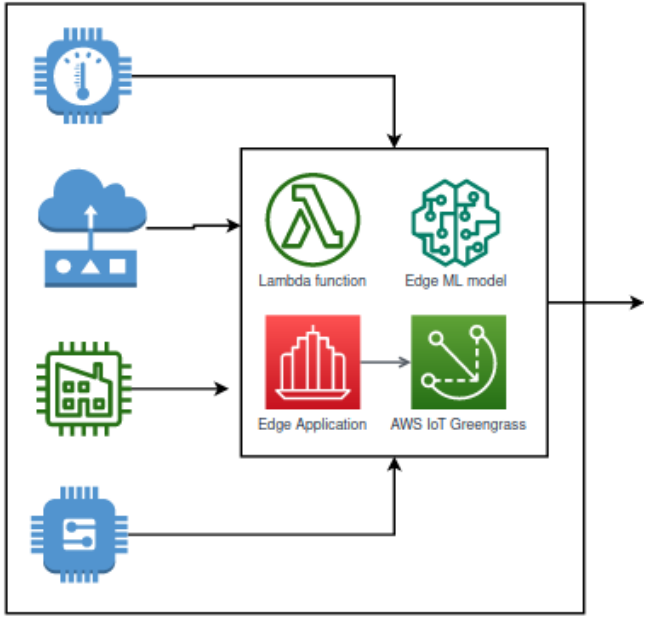

Fig. 3 AWS architectural components of edge layer

## IV. THE CLOUD LAYER

The data generated by IoT devices and relevant batch sources is stored in the cloud for further consumption and archiving. The cloud services provide compute, storage and network infrastructure needed for rigorous data processing of these large volumes of data. The scalable, secure and cost effective cloud infrastructure is leveraged to train, test and evaluate AI/ML models that deliver insights and inferences to businesses. This data is also fed into various business intelligence dashboards and the internal and external applications as needed by the businesses.

The cloud gateway component enables smooth transmission of data between the edge gateway to cloud gateway in a secure and efficient manner. The cloud gateway also facilitates the communication between edge and cloud by ensuring compatibility with needed protocols.

The data received from the edge layer needs to be ingested, buffered and processed both for real time inferences and batch storage. This makes the streaming data available for applications hosted on cloud with minimum delay.

### A. Streaming data processing

On the streaming data in cloud, there will be a need for a data streaming service, a data Extract, Transform, Load (ETL) service and a streaming data analytics service. The data streaming service efficiently captures, processes, and stores data streams. The streaming ETL service effectively captures the data and performs the needed transformation and delivers the transformed data to downstream services such as data lake, data store and analytics services. Streaming analytics component, analyzes the data in real time and provides real time insights into the processes. These insights may help capture anomalies, issues and trends in real time, thus enabling quicker intervention and corrective actions.

The component of streaming ingestion and processing can be divided into hot module and cold module. The hot module processes and provides real time insights from streamed data in the cloud. The hot module also makes the data available to real time databases for needed analytics and processing. The cold module enables long term storage of data. This data is made available for intensive and elaborate batch processing. These may include developing Artificial Intelligence (AI) and Machine Learning (ML) models or dashboarding of data batch data. Fig. 4 shows the architectural components of streaming data processing in hot and cold modules.

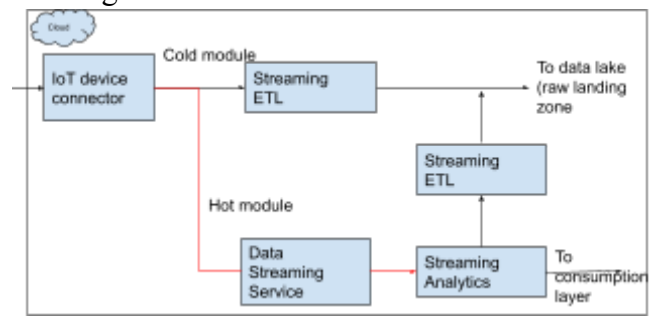

Fig. 4 Reference architecture for streaming data

The Fig. 5 shows the streaming data processing architecture in AWS cloud. The IoT core connects with the edge layer and channels in through Kinesis data stream and Kinesis Firehose. The Kinesis Data Streams stores and ingests streaming data in real time. The Kinesis Data Analytics receives the data and transforms and analyzes the data in real time. The transformed data is sent to the consumption layer for further processing by downstream applications. The transformed data is also ingested into landing raw bucket for long term storage and analysis.

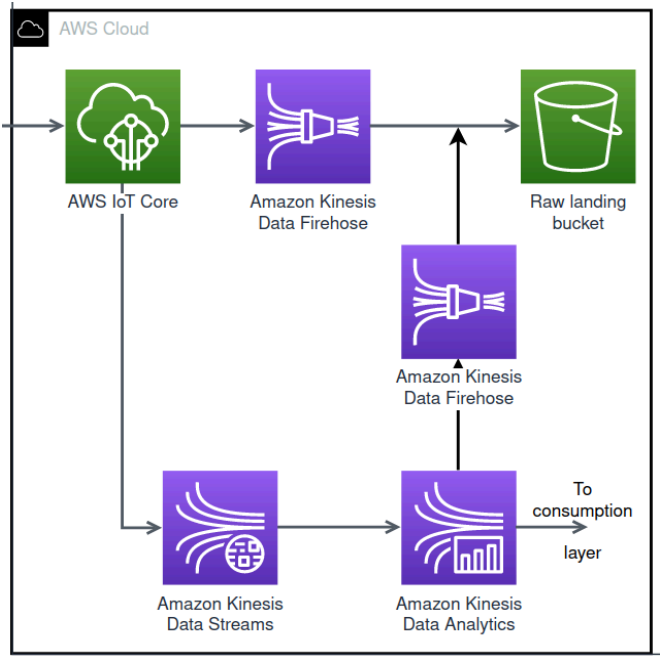

Fig. 5 AWS Reference architecture for streaming data

The batch data in the cloud needs to be received, transformed and stored for efficient use by various business applications and the advanced analytics modules. There are various approaches suggested to achieve this [11-12]. We recommend a zone based modern data strategy based architecture as shown in Fig. 6.

*B. Raw landing zone*

All data that needs to be stored for batch analysis is received in the raw landing zone. The data here available in its raw format. This serves as an optimal node to perform preliminary data quality checks, business compliance and regulatory compliance validations. This is also an optimal node to enforce the data security and governance requirements. Here the masking and encryption needs are taken care of. The organization can identify the sensitive data fields that need to be accessed only by authorized individuals and applications and encrypt or mask it. To further improve the privacy of data, a customized Personally Identifiable Information (PII) detection and removal algorithm can be implemented. This ensures that PII data is identified early in the pipeline and appropriate actions be taken.

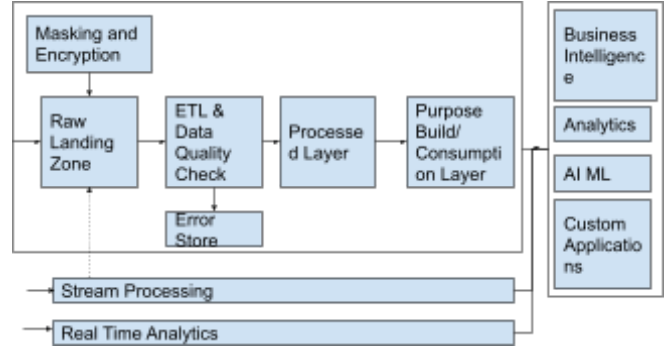

Fig. 6 Reference architecture at cloud layer

The preliminary data quality checks are enforced at this layer. This results in data not meeting the business and quality requirements and is moved to the error store. The data in the error store is evaluated periodically or based on triggers by domain experts. This provides an opportunity to evaluate and find the cause of rejection. Based on the error store analysis findings the data sources can be fixed at the source points or the data acceptance logic is revised. Once the data is processed and made available in the processed layer for downstream consumption, the data from this zone is archived and finally removed.

*C. Processed zone*

The processed layer containing the data is sourced from the landing zone. In this zone the data is stored for long term usage. This serves as a single source of trusted data for downstream processes and applications. The data in this zone is processed to meet the overall business needs. This results in the data that is enriched, indexed data with relevant metadata.

*D. Purpose built zone*

The purpose built layer sources data from the processed layer. The purpose built layer enables a specific set of business applications and advanced analytics. This enables reusable data sets that can be sourced by various applications and teams. The second phase of compliance and quality checks are enforced to ensure that the high quality data is made available to business applications and teams.

*E. Data consumption layer*

The consumption layer has the tools and services needed by data consumer groups. These may include Business Intelligence (BI) developers who use the BI dashboards that are consumed by business end users. These also include machine learning platforms where data scientists use the data from a purpose built layer or the processed layer to build machine learning models. Further the data in the purpose built layers can also be made available to any internal business application or appropriate third party application.

The Fig. 7 shows a holistic modern data architecture in AWS cloud.

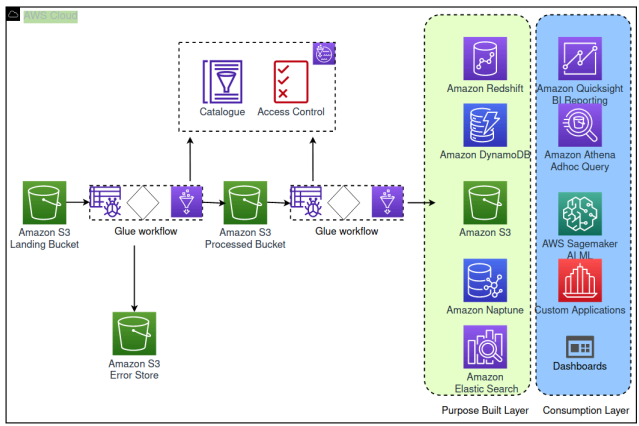

Fig. 7 Reference architecture for modern data strategy

*F. Data governance*

A key component of successful cloud data strategy is effective governance. Data governance ensures that the high quality and trustworthy data is secure, private and easily available to authorized persons. This is a well recognized need in both industry and academia. The governance aspects are addressed in detail in various publications [13-14]. The data lake stores a large amount of data and goes through various transformations between intermediary layers. These sources, transformations and lineages need to be stored and tracked in a metadata repository.

V. CLOUD TO EDGE MACHINE LEARNING MODEL DEPLOYMENT

Data scientists, machine learning engineers and analytics professionals access the data and develop Machine Learning (ML) and Artificial Intelligence (AI) applications that add value to business. Often, the machine learning models developed in the cloud need to be deployed at the edge layer. The machine learning model development process is well established for batch data and cloud or on-premise deployment. The model development process starts by formulating the business problem into data science problem and sourcing and preparing the needed data for model building. This step also involves feature augmentation as needed by use case. This is followed by thorough exploratory analysis to obtain insights into data and evaluating various approaches to algorithm selection, model training, testing and validation. Once the model meets the desired goal the model is taken up for deployment. The ML model building process is a well studied topic and various notable studies include [15-19].

Machine learning on edge brings several challenges that need to be addressed. The edge devices have relatively lower storage and computation power. In addition they operate under various device related constraints. So one needs to optimize the model for the target device at edge to reduce run time without sacrificing accuracy. The ML algorithms should also be power efficient. In some cases, the model inference needs to get data from multiple IoT devices, due consideration should be given for distributed learning algorithms.

Fig. 8 shows the cloud to edge model deployment architecture.

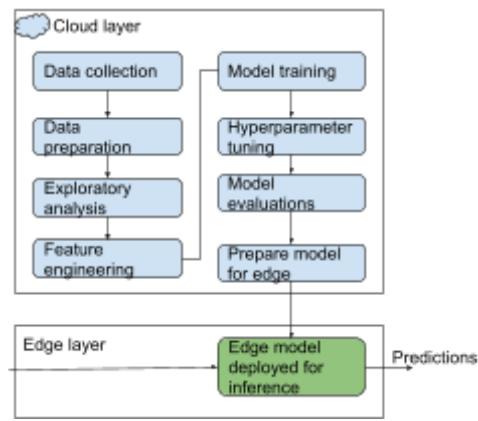

Fig. 8 Architecture for ML model deployment cloud to edge

The AWS implementation of this architecture is shown in Fig. 9. The data needed for model building is sourced from a processed S3 bucket. The SageMaker Data Wrangler is used to prepare the data for model building. The model is developed using SageMaker. The SageMaker Neo is used to optimize the model for edge devices. The model is then deployed to edge for real time inferences.

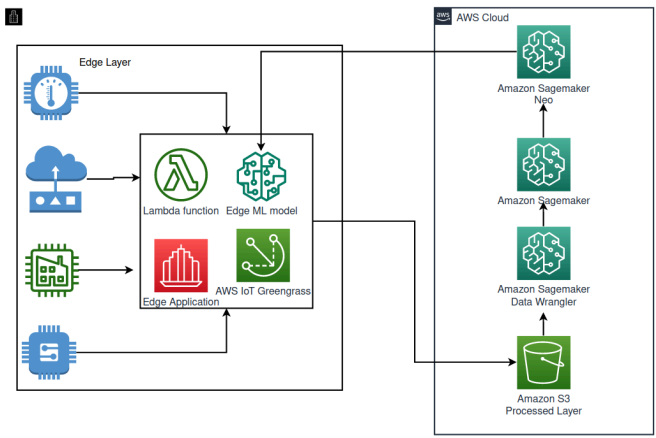

Fig. 9 AWS Architecture for ML model deployment cloud to edge

## VI. Conclusions

The paper provided a secure and efficient end to end edge to cloud data and analytics strategy. The patterns and reference architectures for the device layer, edge layer and the cloud layer were provided. The paper also discussed the machine learning model deployment from cloud to edge. The practical implementation of the reference architectures in cloud were also discussed.